\documentclass[12pt]{article}

\usepackage{amsmath}
\usepackage{amssymb}
\usepackage{epsf}

\newcommand{\eq}{$$} 
\newcommand{\beq}{\begin{equation}} 
\newcommand{\eeq}{\end{equation}\noindent}
\newcommand{\bear}{\begin{eqnarray*}}
\newcommand{\eear}{\end{eqnarray*}\noindent}
\newcommand{\bearn}{\begin{eqnarray}}
\newcommand{\eearn}{\end{eqnarray}\noindent}

\makeatletter 
\def\section{
\@startsection {section}{1}{\z@}{-3.5ex plus -1ex minus 
 -.2ex}{2.3ex plus .2ex}{\large\bf}}

\def\subsection{\@startsection{subsection}{2}{\z@}{-3.25ex plus -1ex minus 
 -.2ex}{1.5ex plus .2ex}{\normalsize\bf}}

\def\subsubsection{\@startsection{subsubsection}{3}{\z@}{-3.25ex plus
 -1ex minus -.2ex}{1.5ex plus .2ex}{\normalsize}}
\makeatother   

\begin{document}

\title{
\normalsize{\bf DISORIENTED CHIRAL CONDENSATE\\FORMATION IN\\HEAVY ION COLLISIONS?}
\thanks{Talk at TH-2002, {\it 'International Conference on 
Theoretical Physics'}, Paris, France.}}

\author{
Julien Serreau\thanks{email: serreau@thphys.uni-heidelberg.de}
\\
[1.ex]\normalsize{Institut f{\"u}r Theoretische Physik der 
Universit{\"a}t Heidelberg}\\
\normalsize{Philosophenweg 16, 69120 Heidelberg, Germany}
}

\date{
\normalsize{February 17, 2003}\\
\normalsize{HD-THEP-03-08}}

\begin{titlepage}
\maketitle
\def\thepage{}          

\vspace{-.5cm}
\begin{center}
\normalsize{\it Dedicated to Andr\'e Krzywicki}
\end{center}

\begin{abstract}
One of the main aims of present and upcoming high energy heavy ion collision 
experiments is to study new phases of matter at extreme temperature 
and density. It is expected that a nontrivial classical pion field 
configuration can occasionally form during the out-of-equilibrium chiral 
phase transition. We have recently shown that, contrarily to what has been 
assumed so far, this configuration is not identical to the so-called 
disoriented chiral condensate (DCC), proposed in the early 1990's. A 
detailed analysis reveals that a more realistic picture is that of an 
``unpolarized'' DCC, where the Fourier modes of the field have completely 
independent orientations in isospin space instead of being aligned with 
each other as in the original DCC. This has important implications 
concerning the possible detection of the phenomenon. In particular, 
the main expected signature of the original DCC, which is used in 
most experimental searches, is absent in the unpolarized case. We 
point out that the fact that no evidence of DCC formation has been
reported so far in nuclear collisions actually agrees with our 
present theoretical understanding. New experimental strategies should 
be designed to look for the unpolarized DCC in existing data from SPS 
as well as in future searches at RHIC and LHC.
\end{abstract}

\end{titlepage}

\section{The Disoriented Chiral Condensate}

Large number of pions with energies below or of the order of a few hundred
MeV are commonly produced in high energy hadronic or nuclear collisions
experiments. Very soon, it has been proposed to interpret this phenomenon as 
resulting from classical radiation, in analogy with the classical
nature of electromagnetic waves corresponding to a large number of 
photons. This idea has received only a marginal attention until the early 
$1990$'s where it has been rediscovered and further developed, on solid 
theoretical grounds~\cite{Blaizot:1996js,Blaizot:1992at,Bjorken:1991sg}. 
In Ref.~\cite{Blaizot:1992at}, Blaizot and Krzywicki studied the dynamics of 
a classical pion field configuration described by the non-linear sigma model, 
in the context of a simple idealized model of a high energy collision. They 
found a solution where the classical pion field oscillates coherently in a 
given direction in isospin space. In parallel, Bjorken proposed that, in such 
collisions, the chiral quark condensate could be momentarily misaligned from 
its vacuum orientation in isospin space, resulting in a non-trivial long 
wavelength configuration of the pion field~\cite{Bjorken:1991sg}. This 
intuitive picture corresponds to Blaizot and Krzywicki's solution. The 
``disoriented chiral condensate'' (DCC) subsequently decays toward ordinary 
vacuum by coherent (classical) radiation of soft pions. The emitted 
pions can be though of as being in a coherent state~\cite{Amelino-Camelia:1997in}, 
schematically\footnote{Strictly speaking, the following state contains
components of arbitrary charge. Generalized coherent or squeezed states of
definite charge can be constructed~\cite{coherent}. This is, however, of
no relevance for the present discussion.}:r
\begin{equation}
 | {\rm DCC} \rangle \sim \exp \left(\sum_{i=1}^3\int d^3k \,\,
 {\rm J}_a ({\bf k})\, {\rm a}_a^\dagger({\bf k}) \right)|0\rangle \, ,
\label{DCCstate}
\end{equation}
where~${\rm a}_a^\dagger({\bf k})$ is the creation operator of a free pion
with isospin~$a=1,2,3$ and momentum~${\bf k}$ and where ${\rm J}_a ({\bf k})$
is the classical source emitting pions (summation over repeated indices is
implied). The latter has the following form:
\begin{equation}
 {\rm J}_a ({\bf k}) = {\rm J} ({\bf k}) \, {\rm e}_a \, ,
\label{polarized}
\end{equation}
that is, its orientation in isospin space is {\em momentum independent}.
This reflects the fact that the disorientation of the pion field 
in isospin space is assumed to be constant in space (in the volume
occupied by the DCC).
The classical pion field configuration corresponding to this state is 
therefore a superposition of waves, which oscillate in the very 
same direction in isospin space. One can speak of a (linearly) polarized 
field configuration~\cite{Serreau:2000tb}. This property is at the origin 
of the most striking signature of the DCC, namely the anomalously large 
event-by-event fluctuations of the neutral fraction of the {\em total} number 
of radiated pions
$$
 f=\frac{N_{\pi^0}}{N_{\pi^0}+N_{\pi^\pm}}
 \equiv \frac{N_{\pi^3}}{N_{\pi^1}+N_{\pi^2}+N_{\pi^3}} \, ,
$$
where $N_{\pi^a}\equiv \int d^3k \, |{\rm J}_a (\bf k)|^2$ is the number of 
pions of type $a$. Indeed, exploiting isospin symmetry of strong
interactions, it is an easy exercise to show that the probability distribution
of the above ratio is given by the inverse square root law (see 
e.g.~\cite{Blaizot:1992at,Bjorken:1991sg}):
\begin{equation}
 P(f) = \frac{1}{2\sqrt f} \, .
\label{neutralfluct}
\end{equation}
This is a direct manifestation of the coherence of the DCC state and is to 
be contrasted with the sharply peaked binomial distribution one obtains in 
the case of incoherent pion production. In particular, 
the probability that less than $10 \%$ of the total number of radiated 
pions\footnote{Of course, this only concerns the pions emitted from the DCC.} 
be neutral is expected to be as large as $30 \%$. This striking property is
at the basis of all present strategies for experimental searches (see 
e.g.~\cite{Mohanty:2002ue} and references therein)\footnote{Various other 
possible signatures have also been discussed~\cite{Huang:1994xu}.}.

\section{A dynamical scenario: the out-of-equilibrium chiral phase transition}

Although it had been understood that the DCC field configuration is a  
solution of the low energy dynamics of strong interactions, a microscopic 
understanding of its possible formation was still lacking. An important progress 
has been made in this respect in 1993, when K.~Rajagopal and F.~Wilczek 
realized that a classical long wavelength pion field configuration could 
indeed be formed during the out of equilibrium chiral phase transition in 
heavy ion collisions: the rapid expansion of the system results in a sudden 
suppression of initial fluctuations (quenching) which in turn triggers a 
dramatic amplification of soft pion modes~\cite{Rajagopal:1993ah}.

\subsection{\label{sec:quench}The quench scenario}

To model the dynamics of the far from equilibrium chiral phase transition, 
Rajagopal and Wilczek considered the classical $O(4)$ linear sigma model, 
with action
\begin{equation}
 \mathcal S = \int d^4x \, 
 \left\{ \frac{1}{2} \, \partial_{\mu} \phi_a \,\partial^{\mu} \phi_a -
	\frac{\lambda}{4} \, \Big(\phi_a \phi_a - v^2 \Big)^2 + 
	H \sigma \right\} \, .
 \label{action}
\end{equation}
The parameters $v$, $\lambda$ and $H$ are 
related to physical quantities via:
\eq
 m_\pi^2 = m_\sigma^2 - 2 \lambda f_\pi^2 =
 \lambda \left( f_\pi^2 - v^2 \right) \, , \, \,
 \mbox{and} \, \, H = f_\pi m_\pi^2 \, .
\eq
and are chosen so that $m_\pi=135$~MeV, $m_\sigma=600$~MeV and $f_\pi=92.5$~MeV.
The corresponding equations of motion are evolved numerically on a lattice for 
a given initial field configuration. The dramatic effect of expansion is modeled 
by assuming an instantaneous quench from above to below the critical temperature
at initial time. In practice, the values of the field $\phi_a$ and its first 
time derivative $\dot \phi_a$ are chosen as independent random variables at 
each site of a lattice\footnote{The lattice spacing has therefore the physical 
meaning of the correlation length in the high temperature phase.}. They are 
sampled from Gaussian distributions centered at $\phi_a = \dot \phi_a = 0$,
corresponding to a high temperature symmetric phase, and with variances
(in lattice units) $\langle \phi_a^2 \rangle = v^2/16$ and 
$\langle \dot \phi_a^2 \rangle = v^2/4$, which are {\em smaller} 
than what they would be in the high temperature phase, thereby 
describing quenched fluctuations.

\begin{figure}[h]
 \centering
 \begin{minipage}[c]{6.cm}
  \caption{\label{fig:rajwil}\small \it Time evolution of the 
  power in the Fourier modes of the pion field $\pi_3$ for quenched 
  initial conditions, as in Ref.\ \cite{Rajagopal:1993ah}. Low momentum modes 
  experience dramatic amplification, the softer the mode the larger the 
  amplification, and exhibit coherent oscillations with period $\sim 2\pi/m_\pi$.}
 \end{minipage}
 \hspace{.1cm}
 \begin{minipage}[c]{7.cm}
  \centerline{\epsfxsize=7.cm\epsfbox{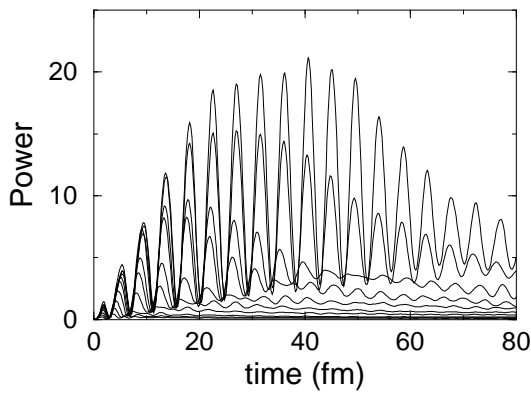}}
 \end{minipage}
\end{figure}

The main result of Ref.\ \cite{Rajagopal:1993ah} is reproduced in 
Fig.\ \ref{fig:rajwil}, which shows the squared amplitude of different field 
modes as a function of time: due to the far from equilibrium initial condition,
one observes a dramatic amplification of low momentum modes of the pion 
field at intermediate times. This phenomenon is analogous to that of domain 
formation after quenching a ferromagnet, or to the growth of quantum fluctuation during the preheating of the early universe after a period of inflation. It can 
be qualitatively understood in a mean-field
approximation~\cite{Rajagopal:1993ah,Mrowczynski:1995at}: at early 
times, the local curvature of the effective potential (effective mass squared) 
is negative and soft modes, for which the associated effective frequency 
is imaginary, experience amplification. This is the so-called spinodal 
instability, well known in condensed matter or nuclear physics. At 
intermediate times, the \mbox{(quasi-)}periodic oscillations of field 
averaged value around the minimum of its effective potential trigger a 
further amplification via the mechanism of parametric
resonance~\cite{Mrowczynski:1995at,Dumitru:2000qy}.

In terms of the underlying quantum dynamics, the amplification of Fourier modes
amplitudes is to be interpreted as the formation of a classical configuration of 
the pion field. Because of expansion, which eventually causes interactions to 
freeze out, the system may be left in such a configuration and subsequently
decay through coherent pion emission.

\subsection{Further developments}

This proposal has attracted much attention and the quench scenario has been 
extensively studied and further developed. In particular, quantum effects in 
the mean field approximation have been included (see e.g.\ 
\cite{Boyanovsky:1994yk,Cooper:ji}) and the drastic quench approximation has 
been abandoned in favor of a description taking explicitely expansion into 
account~\cite{Cooper:ji,Bialas:1994du}. In this context, the importance of 
initial conditions has been pointed out~\cite{Bialas:1994du}. It has been 
realized that, in most cases, initial fluctuations are not suppressed 
efficiently enough and expansion hardly drive the system into a region 
of instability. This observation led to the conclusion that the 
formation of a strong pion field is a rare phenomenon. Exploiting the
assumption of local equilibrium at initial time, it has been possible to 
estimate the probability that a potentially observable signal is produced 
in this scenario~\cite{Krzywicki:1998sc} (see also Ref.\ \cite{Serreau:2000tb}). 
For central Pb-Pb collisions at CERN SPS energies, this analysis yields the 
upper bound
\begin{equation}
 {\rm Proba \, (``observable" \, DCC)} \lesssim 10^{-3} \, .
\label{DCCproba}
\end{equation}
where ``observable'' means that the number of produced DCC pions is required to 
be greater than a typical multiplicity fluctuation~\cite{Krzywicki:1998sc}.
Note that this estimate is obtained within the quench scenario and, therefore, 
corresponds to a conditional probability, which assumes that the conditions 
leading to a quench are met. Thus the DCC is expected to be a fairly rare 
phenomenon, a fact which has important implications. For instance, it 
is very unlikely that more than one ``observable'' DCC domain be formed in a 
single collision\footnote{This does not exclude the possibility that many
DCCs with small number of pions be formed. However, these do not correspond
to classical field configurations.}. Moreover, it is clear that the 
phenomenon should be looked for on an event-by-event basis (see also~\cite{Biro:1997va}). 

The quench scenario has been widely accepted as a microscopic description of 
DCC formation in heavy ion collisions and, as such, has been extensively 
used for DCC phenomenology~\cite{Petersen:1999jc}. Paradoxically, the 
expected large event-by-event fluctuations of the neutral ratio (cf. Eq.\ 
(\ref{neutralfluct})) have never been observed in actual simulations of 
the out-of-equilibrium phase transition~\cite{Gavin:1993bs}.
Of course, deviations from the perfect law (\ref{neutralfluct}) were 
expected on the basis of a microscopic model underlying the original 
picture of a perfectly polarized DCC, cf. Eqs.\ (\ref{DCCstate}) and 
(\ref{polarized}). However, the complete absence of large 
fluctuations in the quench scenario suggests that this idealized 
picture is far from being realized in the context of the 
far-from-equilibrium chiral phase transition~\cite{Serreau:2000tb}.

\section{A more realistic picture: the unpolarized DCC}

Whether the original picture is realized or not in a realistic model is 
a question of great phenomenological importance and requires a careful 
investigation. If the rapid suppression of initial fluctuations provides a 
robust mechanism leading to the formation of a strong coherent pion field, 
it is not clear whether it can explain the hypothetical polarization 
(\ref{polarized}) originally proposed. The key point is to realize that 
the usual assumption of a (locally) thermalized initial state implies 
that the field modes are completely uncorrelated at initial time: they 
are independent (Gaussian) random numbers. Therefore, in order to generate 
a DCC configuration, the microscopic mechanism at work needs not only to 
be efficient in amplifying the modes amplitudes, but it should also 
build correlations between amplified modes.

To study this question, we have performed a detailed statistical analysis 
of generic pion field configurations produced after a quench, with particular 
emphasis on their isospin structure~\cite{Serreau:2000tb}. 
For this purpose, it is sufficient to consider the original model of Rajagopal 
and Wilczek~\cite{Rajagopal:1993ah}, which exhibits all the relevant physical 
features. In particular, the classical field approximation allows one to  
take into account the full nonlinearities, which are crucial for the study of 
correlations. We sample initial field configurations in the Gaussian 
statistical ensemble described in Sec.\ \ref{sec:quench}, and we follow the 
exact time evolution of each configuration by solving numerically the equations 
of motion corresponding to the classical action (\ref{action}). In this way,
we can study the produced configurations on an event-by-event basis. To 
analyze the isospin orientations of distinct field modes, a sensitive 
observable is the neutral fraction of pions in each mode {\bf k},
defined as\footnote{For a more detailed discussion, see 
Ref.~\cite{Serreau:2000tb}.} 
\begin{equation}
 f ({\bf k}) = \frac{n_3({\bf k})}
 {n_1({\bf k})+n_2({\bf k})+n_3({\bf k})} \, ,
\label{ratio}
\end{equation}
where $n_{a=1,\ldots,3} ({\bf k})$ are the averaged occupation 
numbers corresponding to the classical field configuration at the time 
$t_f$ of measurement:
\begin{equation}
 n_a ({\bf k}) = \left| \,
 \frac{i\dot \varphi_a ({\bf k},t_f) + \omega_k \varphi_a ({\bf k},t_f)}
 {\sqrt{2 \omega_k }}
 \, \right|^2 \, .
\label{number}
\end{equation}
Here $\varphi_a$ and $\dot \varphi_a$ denote the Fourier component
of the field and its time derivative respectively and 
$\omega_k=\sqrt{k^2+m_\pi^2}$ is the free pion dispersion relation.
	
In the quench approximation, where the initial fluctuations are suppressed
by hand, the system is initially unstable and all initial configurations 
undergo a dramatic amplification, as in Fig.\ \ref{fig:rajwil}. This procedure
therefore selects the interesting events. To distinguish between different 
mechanisms responsible for amplification, we analyze the properties 
of the statistical ensemble at various final times $t_f$. The plots presented 
here are obtained for $t_f=10$~fm, corresponding to the end of the spinodal
instability, but the results to be discussed below are found to be insensitive 
to this choice. 
\begin{figure}[h]
 \centering
 \begin{minipage}[c]{6.5cm}
  \centerline{\epsfxsize=6.5cm\epsfbox{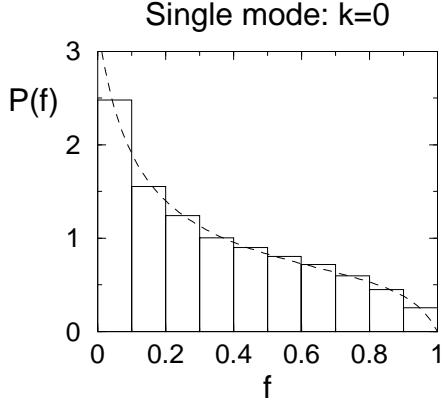}}
 \end{minipage}
 \hspace{.1cm}
 \begin{minipage}[c]{6.cm}
  \caption{\label{fig:mode}\small \it Neutral fraction distribution in a single
  mode, here ${\bf k}=0$, at initial time (dashed line) and at final time,
  here $t_f=10$~fm (histogram).}
 \end{minipage}
\end{figure}

Let us concentrate on the amplified long wavelength modes, which are the 
one we are interested in. The event-by-event distribution of the neutral 
fraction (\ref{ratio}) for the zero mode (the most amplified) is shown in 
Fig.\ \ref{fig:mode}. At a given time, all the amplified modes exhibit a 
similar distribution. In particular, we observe that nothing significant 
happens concerning their isospin structure: the neutral fraction distribution 
is essentially the same as the corresponding one in the initial ensemble.
The latter can be computed exactly~\cite{Serreau:2000tb} and is represented 
by the dashed line in Fig.\ \ref{fig:mode}. For a given mode ${\bf k}$, it
reads\footnote{There is a subtlety associated with the choice 
of boundary conditions for the field~\cite{Serreau:2000tb}. The following formula 
is obtained for Neumann boundary conditions, which are convenient for discussing 
the question of polarization.}
\begin{equation}
 P (f) = \frac{1}{2} \, \Big[ F_\Omega (f) + F_{-\Omega} (f) \Big] \, ,
 \label{indist}
\end{equation}
where
$$
 F_\Omega (f) = [ \Omega - (1-f) ] 
 \left( \frac{\Omega + 1}{\Omega - (1-2f)} \right)^{3/2} \, ,
$$
and\footnote{For a general Gaussian ensemble, one has 
$\Omega = (b^2 + \omega_k a^2)/(b^2 - \omega_k a^2)$,
where $a^2\equiv\langle\varphi_a^2\rangle$ and 
$b^2\equiv\langle\dot\varphi_a^2\rangle$ are the variances of 
the Fourier components of the field and its time derivative 
respectively.}
$$
 \Omega = \frac{4 + \omega_k^2}{4 - \omega_k^2} \, .
$$
Although not exactly~$1/\sqrt f$, the distribution is very broad,
exhibiting large fluctuations around the mean value $\langle f\rangle=1/3$, 
which is the relevant point for phenomenology. As emphasized above, 
these fluctuations are a direct manifestation of the classical 
nature of the field modes. The small deviations from the perfect law
(\ref{polarized}) come from the fact that the latter are not strictly 
linearly polarized~\cite{Serreau:2000tb}.

\begin{figure}[h]
 \centering
 \begin{minipage}[c]{6.cm}
  \caption{\label{fig:correl}\small \it Normalized connected correlation 
   between the neutral fraction in the zero mode and in mode k as a function of 
   momentum.}
 \end{minipage}
 \hspace{.5cm}
 \begin{minipage}[c]{6.8cm}
  \centerline{\epsfxsize=6.5cm\epsfbox{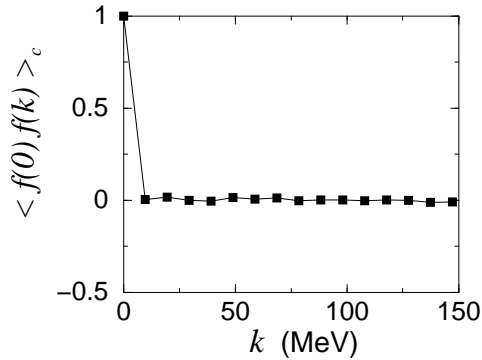}}
 \end{minipage}
\end{figure}

\begin{figure}[h]
 \centering
 \begin{minipage}[c]{6.5cm}
  \centerline{\epsfxsize=6.5cm\epsfbox{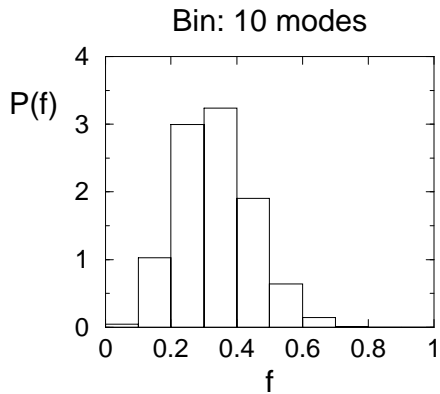}}
 \end{minipage}
 \hspace{.1cm}
 \begin{minipage}[c]{6.cm}
  \caption{\label{fig:bin}\small \it Event-by-event distribution of the 
  neutral fraction in a bin in momentum space containing 10 amplified
  modes. The expected DCC signal is already considerably reduced, due 
  to the absence of correlation between modes.}
 \end{minipage}
\end{figure}

However, we find that the that distinct modes have completely independent 
polarizations in isospin space, as can be seen on Fig.\ \ref{fig:correl},
which shows the correlation between the neutral fractions in different 
(amplified) modes. These modes oscillate coherently, but their 
directions of oscillation in isospin space are completely random. 
In other words, different modes act as independent DCCs.
This has the important phenomenological consequence that the large 
event-by-event fluctuations of the neutral fraction are rapidly washed 
out when the contributions of several modes are added in a momentum bin 
-- as commonly done in experimental analysis -- even when one limits one's 
attention to soft modes only. This is illustrated on Fig.\ \ref{fig:bin}, 
where we show the neutral fraction distribution in a bin containing 
$10$ modes. The expected signal is considerably reduced, already for 
such a small bin. This explains the absence of large fluctuations
reported in previous works~\cite{Gavin:1993bs}, where the authors
typically considered the contribution of a large number of 
modes\footnote{The absence of large fluctuations has also been 
reported in a recent work in a slightly different 
context~\cite{Holzwarth:2002wv}.}.

\begin{figure}[h]
\centerline{\epsfxsize=8.5cm\epsfbox{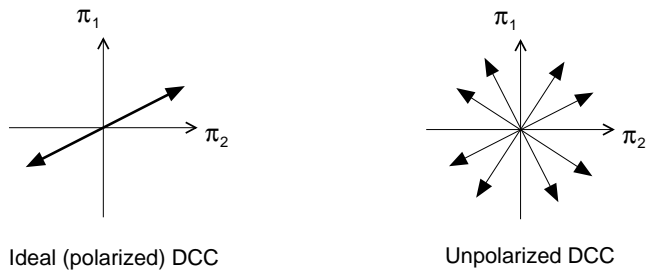}}
\caption{\label{fig:DCCs}\small \it Schematic representation of the 
ideal DCC configuration (left), where all field modes oscillate in the 
same direction in isospin space (see Eq.\ (\ref{polarized})), and of the 
unpolarized DCC formed after a quench (right), where the modes have 
independent directions of oscillation.}
\end{figure}

In conclusion, contrarily to what is usually assumed, the 
non-linear dynamics does not build the required correlation 
between modes: the state produced in the simplest form of the quench 
scenario, where no correlations are present initially, is not  
the originally proposed DCC. Our detailed analysis reveals that a more
realistic picture is that of a superposition of waves having independent 
orientations in isospin space: an ``unpolarized'' DCC configuration, 
as depicted on Fig.\ \ref{fig:DCCs}. 

Of course, one cannot exclude the possibility that the required correlations 
between modes be indeed (partially) formed in actual nuclear or hadronic 
collisions by means of some other mechanism\footnote{For example, the scenario 
proposed in Ref.\ \cite{Asakawa:1998st} could give rise to correlations in the
initial state.}. The resulting configuration would lie between the two extreme 
pictures represented on Fig.\ \ref{fig:DCCs}. Our point here is that there exits, 
at present, no microscopic justification for the polarized DCC state which has 
been looked for so far in experiments~\cite{Brooks:1999xy,Aggarwal:1997hd}. 
Instead, within our present theoretical understanding of DCC formation in heavy 
ion collisions, one expects an unpolarized state to be produced. It is, therefore, 
important to take this information into account in further theoretical
as well as experimental investigations. 

\section{What next?}

Since it has been proposed in the early 1990's, the idea that a DCC might be 
formed in high energy collisions experiments has triggered an intense activity,
both theoretical and experimental. Attempts to detect this phenomenon include
the T-864 MiniMax experiment, a dedicated search in proton-antiproton collisions 
at the Fermilab Tevatron~\cite{Brooks:1999xy}, or event-by-event analysis of Pb-Pb 
collisions by the WA98 and NA49 collaborations at CERN SPS~\cite{Aggarwal:1997hd}. 
No clear experimental evidence for DCC formation has been reported
so far\footnote{DCC has, however, been proposed to be responsible for so-called
Centauro events, reported in the cosmic ray literature~\cite{Lattes:1980wk,Blaizot:1996js}.}.
This is a bit disappointing in view of the fact that the phenomenon seems so
natural, at least in the context of heavy ion collisions. Indeed, the growth of 
long wavelength fluctuations after a quench is a generic phenomenon, observed 
in many actual situations, for example in condensed matter experiments, and which 
is likely to happen in heavy ion collisions as well. In particular, it has been 
shown to occur in various different models~\cite{Bedaque:fa,Dumitru:2000qy}.
The absence of experimental evidence may indicate that the appropriate conditions 
for DCC formation have not been met yet (see the interesting discussion in 
Ref.~\cite{Rajagopal:2000yt}, where it is argued that semi-peripheral heavy 
ion collisions at RHIC energies might provide more favorable conditions for 
efficient quenching).

However, we would like to point out that the present experimental 
situation can actually be simply understood within our present theoretical 
understanding. Indeed, the upper bound (\ref{DCCproba}) for the probability 
of DCC formation at SPS energies is at the edge of the current experimental 
limit probed by the WA98 and NA49 collaborations~\cite{Aggarwal:1997hd}. 
Moreover, our results concerning the (iso-)polarization of the DCC indicates 
that the phenomenon we are seeking may actually be more difficult to observe 
than originally thought. In particular, it would be of first interest to 
analyze existing data in the light of the present results and to design new 
strategies for future DCC searches at RHIC and LHC~\cite{Gladysz-Dziadus:2002ps}, 
taking into account its unpolarized nature.

Besides providing a useful tool for the experimental study of the QCD chiral 
phase transition~\cite{Rajagopal:2000yt}, the detection of a DCC is 
of fundamental interest on its own. Indeed, the possible formation of
a coherent classical pion field is to be expected on very general grounds
as a direct consequence of the bosonic nature of pions. This phenomenon
is analogous to that of Bose-Einstein condensation, currently observed
in condensed matter experiments. The fact that such a macroscopic 
wave of QCD matter may be produced is a very exciting possibility. 
I think it is worth the effort.

\section*{Acknowledgments}
I acknowledge a fruitful collaboration with A.~Krzywicki about 
the present subject as well as many interesting conversations about
physics and other topics. I thank J.-P.~Blaizot 
for stimulating discussions.


\begin{thebibliography}{99}

\bibitem{Blaizot:1996js}
See the review by
J.~P.~Blaizot and A.~Krzywicki,
Acta Phys.\ Polon.\ {\bf B27} (1996) 1687 and references therein.

\bibitem{Blaizot:1992at}
J.~P.~Blaizot, A.~Krzywicki,
Phys.\ Rev.\ {\bf D46} (1992) 246;
Phys.\ Rev.\ {\bf D50} (1994) 442.

\bibitem{Bjorken:1991sg}
J.~D.~Bjorken,
SLAC-PUB-5673;
Acta Phys.\ Polon.\ {\bf B23} (1992) 637.

\bibitem{Amelino-Camelia:1997in}
G.~Amelino-Camelia, J.~D.~Bjorken, S.~E.~Larsson,
Phys.\ Rev.\ {\bf D56} (1997) 6942.

\bibitem{coherent}
D.~Horn, R.~Silver, Annals of Phys. {\bf 66} (1971) 509;
J.~C.~Botke, D.~J.~Scalapino, R.~L.~Sugar,
Phys.\ Rev.\ {\bf D9} (1974) 813.

\bibitem{Serreau:2000tb}
J.~Serreau,
Phys.\ Rev.\ {\bf D63} (2001) 054003; see also
J.~Serreau,
PhD Thesis, LPT-ORSAY-01-27,
[hep-ph/0104023].

\bibitem{Mohanty:2002ue}
B.~Mohanty, D.~P.~Mahapatra, T.~K.~Nayak,
Phys.\ Rev.\ {\bf C66} (2002) 044901;
B.~Mohanty, T.~K.~Nayak, D.~P.~Mahapatra, Y.~P.~Viyogi,
\mbox{nucl-ex/0211007.}

\bibitem{Huang:1994xu}
Z.~Huang, M.~Suzuki, X.~N.~Wang,
Phys.\ Rev.\ {\bf D50} (1994) 2277;
D.~Boyanovsky, H.~J.~de Vega, R.~Holman, S.~Prem Kumar,
Phys.\ Rev.\ {\bf D56} (1997) 5233;
Y.~Kluger, V.~Koch, J.~Randrup, X.~N.~Wang,
Phys.\ Rev.\ {\bf C57} (1998) 280;
J.~I.~Kapusta, S.~M.~Wong,
Phys.\ Rev.\ Lett.\  {\bf 86} (2001) 4251;
R.~C.~Hwa, C.~B.~Yang,
Phys.\ Lett.\ {\bf B534} (2002) 69.

\bibitem{Rajagopal:1993ah}
K.~Rajagopal, F.~Wilczek,
Nucl.\ Phys.\ {\bf B404} (1993) 577.

\bibitem{Mrowczynski:1995at}
S.~Mr\'owczy\'nski, B.~M\"uller,
Phys.\ Lett.\ {\bf B363} (1995) 1;
D.~I.~Kaiser,
Phys.\ Rev.\ {\bf D59} (1999) 117901.

\bibitem{Dumitru:2000qy}
A.~Dumitru, O.~Scavenius,
Phys.\ Rev.\ {\bf D62} (2000) 076004.

\bibitem{Boyanovsky:1994yk}
D.~Boyanovsky, H.~J.~de Vega, R.~Holman,
Phys.\ Rev.\ {\bf D51} (1995) 734.

\bibitem{Cooper:ji}
F.~Cooper, Y.~Kluger, E.~Mottola, J.~P.~Paz,
Phys.\ Rev.\ {\bf D51} (1995) 2377;
M.~A.~Lampert, J.~F.~Dawson, F.~Cooper,
Phys.\ Rev.\ {\bf D54} (1996) 2213.

\bibitem{Bialas:1994du}
A.~Bialas, W.~Czyz, M.~Gmyrek,
Phys.\ Rev.\ {\bf D51} (1995) 3739;
J.~Randrup,
Phys.\ Rev.\ Lett.\  {\bf 77} (1996) 1226.

\bibitem{Krzywicki:1998sc}
A.~Krzywicki, J.~Serreau,
Phys.\ Lett.\ {\bf B448} (1999) 257.

\bibitem{Biro:1997va}
T.~S.~Bir\'o, C.~Greiner,
Phys.\ Rev.\ Lett.\  {\bf 79} (1997) 3138.

\bibitem{Petersen:1999jc}
See e.g. 
T.~C.~Petersen, J.~Randrup,
Phys.\ Rev.\ {\bf C61} (2000) 024906.

\bibitem{Gavin:1993bs}
S.~Gavin, A.~Gocksch, R.~D.~Pisarski,
Phys.\ Rev.\ Lett.\  {\bf 72} (1994) 2143;
J.~Randrup,
Nucl.\ Phys.\ {\bf A616} (1997) 531;
K.~Rajagopal,
talk given at Hirschegg 97, hep-ph/9703258.

\bibitem{Holzwarth:2002wv}
G.~Holzwarth, J.~Klomfass,
Phys.\ Rev.\ {\bf D66} (2002) 045032.

\bibitem{Asakawa:1998st}
M.~Asakawa, H.~Minakata, B.~M\"uller,
Phys.\ Rev.\ {\bf D58} (1998) 094011.

\bibitem{Brooks:1999xy}
MiniMax Collaboration, T.~C.~Brooks {\it et al.},
Phys.\ Rev.\ {\bf D61} (2000) 032003.

\bibitem{Aggarwal:1997hd}
WA98 Collaboration, M.~M.~Aggarwal {\it et al.},
Phys.\ Lett.\ {\bf B420} (1998) 169; see also
nucl-ex/0206017;
NA49 Collaboration, H.~Appelshauser {\it et al.},
Phys.\ Lett.\ {\bf B459} (1999) 679.

\bibitem{Gladysz-Dziadus:2002ps}
See e.g.\  the event-by-event program of the STAR detector at RHIC, 
\verb"http://www.star.bnl.gov/STAR";
searches of Centauro-like events are planned at LHC e.g.\  with the 
CASTOR detector, see E.~Gladysz-Dziadus {\it et al.},
hep-ex/0209008.

\bibitem{Lattes:1980wk}
C.~M.~Lattes, Y.~Fujimoto, S.~Hasegawa,
Phys.\ Rept.\  {\bf 65} (1980) 151.

\bibitem{Bedaque:fa}
P.~F.~Bedaque, A.~K.~Das,
Mod.\ Phys.\ Lett.\ {\bf A8} (1993) 3151;
A.~Barducci, L.~Caiani, R.~Casalbuoni, M.~Modugno, G.~Pettini, R.~Gatto,
Phys.\ Lett.\ {\bf B369} (1996) 23;
J.~Schaffner-Bielich, J.~Randrup,
Phys.\ Rev.\ {\bf C59} (1999) 3329.

\bibitem{Rajagopal:2000yt}
K.~Rajagopal,
Nucl.\ Phys.\ {\bf A680} (2000) 211.

\end{thebibliography}
\end{document}